\documentclass{elsart5}

\usepackage{graphicx}
\usepackage{epsfig}
\usepackage{amssymb}

\begin{document}

\begin{frontmatter}

\title{Thermal expansion of the quasi two-dimensional magnetic layered compound BaNi$_{2}$V$_{2}$O$_{8}$ under magnetic
fields along $c$-axis.}

\author[aff1,aff2]{W. Knafo\corauthref{cor1}}
\ead{william.knafo@ifp.fzk.de}
\corauth[cor1]{}
\author[aff1]{C. Meingast}
\author[aff1]{K. Grube}
\author[aff1,aff2]{S. Drobnik}
\author[aff1,aff2]{P. Popovich}
\author[aff1]{P. Schweiss}
\author[aff1]{Th. Wolf}
\author[aff1,aff2]{and H. v. L\"{o}hneysen}
\address[aff1]{Forschungszentrum Karlsruhe, Institut f\"{u}r Festk\"{o}rperphysik, D-76021 Karlsruhe, Germany}
\address[aff2]{Physikalisches Institut, Universit\"{a}t Karlsruhe, D-76128 Karlsruhe, Germany}
\received{7 June 2006} \revised{xxx} \accepted{xxx}

\begin{abstract}
The quasi two-dimensional magnetic BaNi$_{2}$V$_{2}$O$_{8}$ is
studied by using high-resolution thermal expansion in magnetic
fields up to 10 T applied along the $c$-axis. A slight increase of
about 1 \% of the three-dimensional antiferromagnetic ordering
temperature $T_N$ is observed at 10 T. Positive and negative
pressure dependencies of $T_N$, respectively, are inferred from
the thermal expansion $\alpha(T)$ for pressures applied along the
$a$- and $c$-axes.

\end{abstract}


\end{frontmatter}

Although the properties of quasi-two dimensional (2D) magnetic
systems have been studied for almost three decades
\cite{dejongh90}, a lot of activity is still devoted to their
study. Of particular relevance is the question of a
Kosterlitz-Thouless description \cite{kosterlitz73} for such
systems, continuing to be an open and challenging debate. One of
the main motivations related to these works comes from the fact
that the high $T_c$ cuprates are also quasi-2D magnetic systems.
Attention focused recently on experimental studies of the layered
compound BaNi$_{2}$V$_{2}$O$_{8}$ \cite{rogado02,heinrich03}. This
system, which is constituted of a honeycomb arrangement of spin-1
ions of Ni$^{2+}$, is characterized by a planar magnetic exchange
responsible for a strongly 2D character. A three-dimensional (3D)
long-range ordering (with the spins aligned in the plane) sets in
below $T_N\simeq48$ K and is explained as a consequence of a tiny,
but non-zero, out-of-plane exchange \cite{rogado02,heinrich03}. An
isotropic, thus Heisenberg-like, susceptibility was also reported
for temperatures above 100 K, and a description within a 2D-XXZ
picture has been proposed to relate the behavior just above $T_N$
to a Kosterlitz-Thouless-like picture
\cite{rogado02,heinrich03,cuccoli03}.\\

To validate or not such a description, an effort has to be made to
determine precisely the role of the exchange parameters (in and
between planes) and of the single-ion anisotropy (planar,
in-plane, etc.), but also the magnetic-field and pressure
dependencies of those quantities.\\

\begin{figure}[b]
    \centering
    \epsfig{file=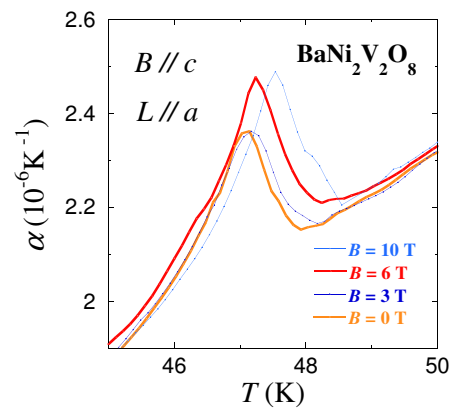,height=53mm}
    \caption{Thermal expansion of BaNi$_{2}$V$_{2}$O$_{8}$ for fields $B$
    applied along $c$, the length change being measured along $a$.}
    \label{fig1}
\end{figure}

We present here a study of the thermal expansion of a single
crystal of BaNi$_{2}$V$_{2}$O$_{8}$ with magnetic fields $B$
applied along the $c$-axis, thus perpendicular to the easy plane.
The sample of about 15 mm$^3$ was grown using the self-flux
method. Thermal expansion was measured using a high-resolution
capacitive dilatometer with a heating rate of 20 mK/s. The length
$L$ along $a$- and $c$-axes was measured with the field applied in
the $c$-direction. The thermal expansion coefficient was then
extracted using $\alpha=1/L*\partial L/\partial
T$.\\

\begin{figure}[b]
    \centering
    \epsfig{file=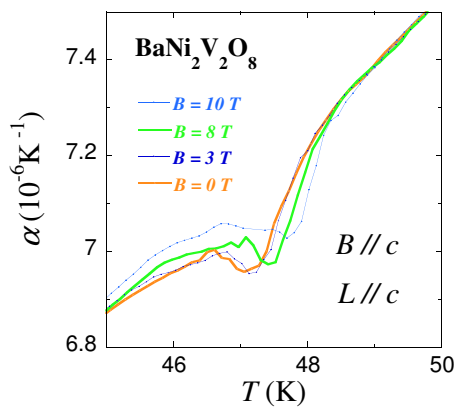,height=53mm}
    \caption{Thermal expansivity of BaNi$_{2}$V$_{2}$O$_{8}$ for fields $B$
    applied along $c$, the length change being measured along $c$.}
    \label{fig2}
\end{figure}

\begin{figure}[t]
    \centering
    \epsfig{file=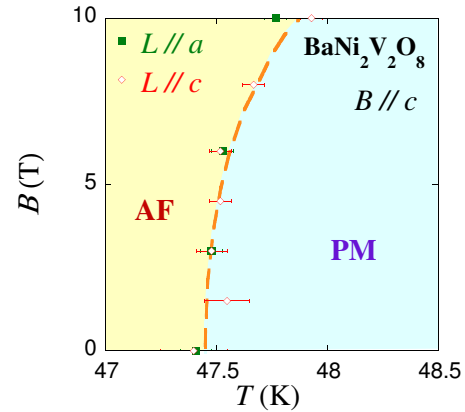,height=54mm}
    \caption{Phase diagram $(B,T)$ of BaNi$_{2}$V$_{2}$O$_{8}$, with
    $B$ applied along $c$. The points come from the two
    thermal-expansion set-up $L$ // $a$ and $L$ // $c$, and the line is a guide to the eyes.}
    \label{fig3}
\end{figure}

In Fig. \ref{fig1}, the temperature variation of the thermal
expansivity obtained in the set-up with $L$ // $a$ is shown for
applied magnetic fields $B=$ 0, 3, 6, and 10 T. A clear step-like
anomaly is observed at $T_N$; in contrast, only a change of slope
was reported previously in the specific heat measured at $B=0$
\cite{rogado02}, which we attribute to insufficient resolution of
the specific-heat data. In the thermal expansion obtained with $L$
// $a$, a positive anomaly signals the antiferromagnetic ordering.
We take for the N\'{e}el temperature the temperature of the
steepest slope of $\alpha(T)$, i.e. the minimum of
$\partial\alpha/\partial T$. Hence $T_N \simeq 47.4$ K at $B=0$.
The application of a magnetic field along the $c$-axis results in
a slight increase of $T_N$ up to 47.8 K at $B=$ 10 T, i.e. of
about 1\%. A small increase of the size of the anomaly is also
obtained for the fields $B=$ 6 and 10 T. Fig. \ref{fig2} shows the
thermal expansion coefficient $\alpha$ obtained in the set-up with
$L$ // $c$ for the magnetic fields $B=$ 0, 3, 8, and 10 T. The
jump associated with $T_N$ is negative and thus of opposite sign
compared to the previous configuration. The N\'{e}el temperature
$T_N$ deduced from this anomaly has a variation with field which
is consistent with that deduced from the transverse measurements,
while the size of the anomaly does not seem to change with field.
The phase diagram derived from these two sets of measurements is
plotted in Fig. \ref{fig3}. Using the Ehrenfest relation, the jump
$\Delta\alpha$ of the thermal expansion at the magnetic ordering
can be related to the pressure dependence of the N\'{e}el
temperature by $\partial T_N/\partial p\propto\Delta\alpha$, where
$p$ is the pressure. Our data thus show a positive pressure
dependence $\partial T_N/\partial p_a>0$ for $p$ applied along $a$
and a negative pressure dependence $\partial T_N/\partial p_c<0$
for $p$ applied along $c$, independently of the magnetic field.
Interpreting these pressure dependencies is not easy, since they
result from the combination of several effects, such as the
pressure dependence of the anisotropy, sensitive to the domain
arrangement, or the pressure dependence of the exchange, and thus
the different super-exchange paths of the system. The increase of
$T_N$ when a magnetic field is applied along the $c$-axis is
associated with an increase of the effective easy-plane
anisotropy, and thus of the XY-character of the spins. Such a
picture was already proposed by Takeda and Koyama to explain the
phase diagram of Mn(HCOO)$_2$2H$_2$O \cite{takeda83}, which was
recently compared to the result of
Monte Carlo simulations by Cuccoli et al. \cite{cuccoli06}.\\

In conclusion, we have performed thermal expansion measurements on
the quasi-2D magnetic system BaNi$_{2}$V$_{2}$O$_{8}$ under
magnetic fieds along the $c$-axis. We have obtained a slight
increase with field of the 3D-ordering temperature $T_N$, while
positive and negative pressure dependencies of $T_N$ are obtained
for $p$ applied along $a$ and $c$, respectively. A forthcoming
article will present a complete study of BaNi$_{2}$V$_{2}$O$_{8}$
using the combination of thermal-expansion, specific-heat, and
magnetization measurements performed with magnetic fields applied
along $c$ and $a$ \cite{knafo06}.\\

We acknowledge useful discussions with L.P. Regnault, C. Boullier,
D. Reznik, T. Roscilde, J. Villain, S. Bayrakci, B. Keimer, and R.
Eder. This work was supported by the Helmholtz-Gemeinschaft
through the Virtual Institute of Research on Quantum Phase
Transitions and Project VH-NG-016.

\end{document}